# Fire detection in a still image using colour information

Oluwarotimi Giwa[1] and Abdsamad Benkrid[2]

School of Engineering
Faculty of Technology
University of Portsmouth
Portsmouth, United Kingdom
`Oluwarotimi.giwa@port.ac.uk`
`abdsamad.benkrid@port.ac.uk`

**Abstract**

Colour analysis is a crucial step in image-based fire detection algorithms. Many of the proposed fire detection algorithms in a still image are prone to false alarms caused by objects with a colour similar to fire. To design a colour-based system with a better false alarm rate, a new colour-differentiating conversion matrix, efficient on images of high colour complexity, is proposed. The elements of this conversion matrix are obtained by performing K-medoids clustering and Particle Swarm Optimisation procedures on a fire sample image with a background of high fire-colour similarity. The proposed conversion matrix is then used to construct two new fire colour detection frameworks. The first detection method is a two-stage non-linear image transformation framework, while the second is a direct transformation of an image with the proposed conversion matrix. A performance comparison of the proposed methods with alternate methods in the literature was carried out. Experimental results indicate that the linear image transformation method outperforms other methods regarding false alarm rate while the non-linear two-stage image transformation method has the best performance on the F-score metric and provides a better trade-off between missed detection and false alarm rate.

## 1 Introduction

Early fire detection in the event of an outbreak is pivotal to prevent loss of life and properties. Traditional smoke detection systems are sensor-based and detect the presence of the by-products of combustion, such as smoke, heat and radiation. As a result, they are not reliable in open space and windy conditions. Moreover, their response speed depends on how quickly the combustion products get close enough to the sensors to activate them. Also, sensors cannot provide sufficient information for proper estimation of the fire's size, location and dynamics. An alternate solution is a computer vision-based detection system, which offers the following advantages: (I) low-cost, early and fast detection (II) visual feedback information about the location and state of fire.

In the literature, three approaches are used to detect fire colour: methods based on statistics or probability distributions, clustering or data mining, and colour space rules. In methods based on colour distribution, a pixel is classified as fire if it belongs to a predetermined colour probability distribution model trained with a set of fire images. For example, Ko et al. [1] modelled each of the RGB colour channels with a unimodal Gaussian distribution and classified a pixel as fire if its overall probability distribution is above a threshold. Chen et al. [2] adopted a hybrid combination of a Gaussian model based on Cb/Cr colour distribution using the YCbCr colour space and a K-means clustering algorithm in the $L^* a^* b^*$ colour space to classify flame pixels in still images. The clustering-based techniques also demonstrated in Chen et al. [2] label all the pixels within a cluster as fire if its centroid is very close to the colour of fire. In another clustering related research, Truong and Kim [3] employed a Fuzzy C-Means Clustering algorithm to detect fire colour in the $L^* a^* b^*$ colour space. The colour space rules [4, 5, 6, 7, 8, 9] establish rules based on the relationships between the colour components of a pixel.

Researchers in image processing have made several attempts to determine the most suitable colour space for fire detection. Chen et al. [4] proposed three rules based on the RGB components. However, the downside of this algorithm is that fire representation in the RGB colour space is not robust against illumination change. Therefore it is not possible to separate a colour between chrominance and intensity. To tackle this limitation, Celik and Demirel [10] transformed the RGB colour space into YCbCr colour space where the separation between luminance and chrominance is possible. Horng et al. [5] represented in the HIS colour space because it simulates the colour sensing properties of the human visual system. Du and Liu [11] carried out a comparative analysis of 18 different colour spaces, using a BoF-based method. They concluded that Srgb and PJF colour spaces are the most effective for flame detection, in terms of classification accuracy and class separability measures. Khatami et al. [12] converted the RGB colour space to a new flame-based colour space in which the fire pixels are highlighted, and the non-fire pixels are dimmed, making the fire-parts in the converted image to be extracted efficiently with the Otsu thresholding algorithm. While this method performs well in a forest fire situation, it is not very efficient in an environment with a higher colour similarity in the background.

This paper introduces a new colour space conversion matrix, more robust against objects or backgrounds of high colour complexity. We adopt the K-medoids clustering and Particle Swarm Optimisation procedures proposed in [12]. However, we use a training fire image with a background of high colour similarity. Also, we propose two new solutions to fire detection, which involve a linear and a non-linear colour space transformation respectively. The rest of this paper is organised as follows: Section 2 discuss the training and fire detection phase of the proposed algorithms. Section 3 presents the experimental results. Section 4 states our conclusion.

# 2 Research Approach

This study aims to obtain a colour space conversion matrix that can differentiate a real fire from an object or background with a high degree of fire-colour similarity, in an image. As a result, we train the conversion matrix with the fire sample shown in Figure 1(a). A feature matrix is constructed from the sample image in Figure 1(b): the pixels from the fire part of the sample image populate the first half of the feature matrix, while the pixels from background occupy the second half. The dimension of each half of the feature matrix is 20 X 40 X 3; the third dimension represents the R G B components. In contrast to [12] - [14], the feature matrix has a low degree of colour dissimilarity between its fire part and non-fire part. In the following sections, we discuss the algorithms to obtain the proposed conversion matrix. Also, we introduce two new fire detection frameworks, which is designed with the conversion matrix.

## 2.1 Evaluating the colour differentiating ability of a conversion matrix

An ideal colour space conversion matrix should be able to highlight all the fire pixels and dim all the non-fire pixels in the feature matrix. According to [12], this colour differentiating ability can be determined by performing a two-class (fire and non-fire pixel) K-medoids clustering procedure on the feature matrix, before and after conversion, using Eq. (2) and Eq. (3) respectively. The conversion process is a linear multiplication of the RGB components of a pixel with the conversion matrix, as expressed in Eq. (1).

$$Y = X * W \tag{1}$$

$$C_1 = \sum_{j=1}^{Z} \sum_{c=1}^{2} |X_j - m_c| \tag{2}$$

$$C_2 = \sum_{j=1}^{Z} \sum_{c=1}^{2} |Y_j - m_c| \tag{3}$$

where $X$ and $Y$ are the intensity values of a pixel before and after conversion respectively; W is a 3 X 3 conversion matrix; $Z$ is the total number of non-medoids pixels; $c$ is the class a pixel belongs to; $m_c$ is a medoid.

The total number of pixels that maintain the same K-medoid class before after conversion can be used to quantify the colour differentiating ability of a conversion matrix. Assuming $X_{ij}$ is a pixel on the original feature matrix at spatial location $(i, j)$ and belongs to a K-medoids cluster $Class(X_{ij})$ (can be a fire or a non-fire class) and $Y_{ij}$ is a converted pixel on the same spatial location that belongs to cluster $Class(Y_{ij})$ (can be a fire or a non-fire class), we express the transformation error as:

$$Cost = \sum_{i=1}^{n} \sum_{j=1}^{n} [\, Class(X_{ij}) \neq Class(Y_{ij})] \tag{4}$$

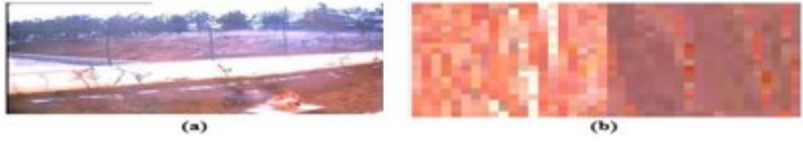

Figure 1: (a) Sample image; (b) feature matrix constructed from the sample image. The fire pixels from the sample image populate the first half of the feature matrix, and non-fire pixels populate the second half.

## 2.2 Obtaining the proper weights of the conversion matrix using Particle Swarm Optimisation algorithm (PSO)

PSO is an intelligent optimisation algorithm that belongs to a class of meta-heuristics optimisation algorithms. Proposed by Eberhart and Kennedy in [15], the algorithm is based on the concept of swarm intelligence and motivated by the social behaviour of animals. The particle with the global best position controls the movement of the whole population. A position is evaluated using a fitness or cost function. On every iteration of the PSO search, each particle keeps a history of its personal best position and replace it with its new position, if its new position is better than its personal best. At the same time, any particle with the overall best position in the population becomes the new global best particle. The PSO implementation in this paper is similar to [12] – [14]: The population is a set of 50 3 X 3 matrixes, initialised with random values. Our goal is to search for an optimum matrix that can transform the feature matrix such that each pixel has the same K-medoid class before and after conversion. During each iteration of our PSO, we update the position or elements of each matrix in the population, using Eq. (5) and Eq. (6). Then, we evaluate the colour differentiating ability of the new positions for each of the matrixes in the population, using the cost function in Eq. (4). We update the personal best of each of the matrix based on the cost of their new positions. If the cost of the new position of a matrix is lower than that of its personal best position, we replace its personal best position with its new position. Also, a matrix with the overall personal best position in the swarm becomes the new global best solution. We terminate the PSO search after 200 iterations or when we obtain a global best solution with a cost value of 0.

$$\overrightarrow{vel_{x,y}}(t+1) = \omega * \overrightarrow{vel_{x,y}}(t) + c_1 r_{1,y}(t)[Pos_{best,x} - \overrightarrow{Pos_{x,y}}(t)] + c_2 r_{2,y}(t)[Glo_{best} - \overrightarrow{Pos_{x,y}}(t)] \tag{5}$$

$$\overrightarrow{Pos_{x,y}}(t+1) = \overrightarrow{Pos_{x,y}}(t) + \overrightarrow{vel_{x,y}}(t) \tag{6}$$

where the vector $\overrightarrow{vel_{x,y}}(t)$ denotes the velocity of matrix $x$ in component $y$ at search iteration $t$; vector $\overrightarrow{Pos_{x,y}}(t)$ represents the position of matrix $x$; $Pos_{best,x}$ is the personal best position of matrix $x$; $Glo_{best}$ is the best fit matrix in the swarm; $c_1$ and $c_2$ are the personal and social (global) acceleration coefficients respectively; $r_{1,y}(t)$ and $r_{2,y}(t)$ are uniformly distributed random numbers between 0 and 1 at time step $t$; $\omega$ is a constriction coefficient which ensures all the particles in the population converge overtime. At the end of the PSO search the value of $Glo_{best}$ is found to be:

$$\begin{bmatrix} 1.7673 & 2.9860 & -0.9186 \\ 0.1479 & -0.9451 & -1.2610 \\ -3.2330 & -2.8938 & -1.3918 \end{bmatrix}$$

## 2.3 Fire detection using a non-linear two-stage image transformation

The first stage of the image transformation is the same as [12]. Firstly, contrast enhancement of the input image is performed according to Eq. (7), to boost the colour-differentiating ability of the conversion matrix. Next, the RGB component of the enhanced image is multiplied by the conversion matrix proposed in [12], using to Eq. (8). Then, Otsu thresholding algorithm is applying on the converted image to remove non-fire pixels, since the fire parts are more highlighted than the non-fire part. At this stage of the algorithm, the system is still very prone to a false alarm, especially on images with a reddish background. Therefore, the image is further processed in stage 2, repeating Stage 1 procedures but using Eq. (9) and Eq. (10).

$$I_c = R^{1.5} G^{0.7} B^{0.9} \tag{7}$$

$$[R_1 \quad G_1 \quad B_1] = [R \quad G \quad B] * \begin{bmatrix} 3.2753 & 1.9701 & 1.8017 \\ -0.0269 & -0.0774 & 0.2938 \\ -3.0439 & -1.9676 & -2.3011 \end{bmatrix} \tag{8}$$

where $I_c$ is the intensity of the constrast enhanced image; $[R \quad G \quad B]$ and $[R_1 \quad G_1 \quad B_1]$ are the colour components of the image before and after conversion respectively.

In Figure 2, we consider an image with a high colour similarity background. Figure 2(a-d) and Figure 2(e-h) illustrate the algorithms of Stage 1 and Stage 2 respectively. Some background pixels are still highlighted as a fire pixel at the end of Stage 1 as depicted in Figure 2(d). Further transformation of the image in stage 2 significantly reduce the number of these background pixels misclassified as fire as shown in Figure 2(h).

$$I_c = R^4 G^{0.9} B^2 \tag{9}$$

$$[R_1 \quad G_1 \quad B_1] = [R \quad G \quad B] * \begin{bmatrix} 1.7673 & 2.9860 & -0.9186 \\ 0.1479 & -0.9451 & -1.2610 \\ -3.2330 & -2.8938 & -1.3918 \end{bmatrix} \tag{10}$$

Figure 3 presents a scenario where the fire parts of the input image are very bright such that the transformation process of stage 1 converts these pixels to white pixels, as demonstrated in Figure 3(a-d). As a result, these fire pixels are eliminated in Stage 2, causing a significant high missed fire detection rate. To resolve this problem, we first label any pixel that is converted to a white pixel as a fire pixel, before proceeding to stage 2, using a threshold of 0.8 on each of the RGB colour channels. Figure 4 presents the flowchart of the non-linear two-stage image transformation algorithm.

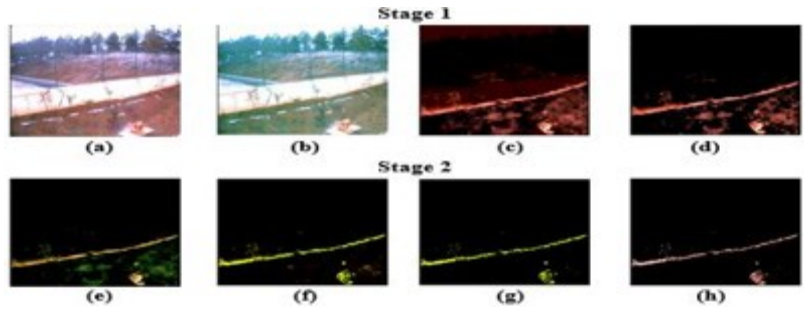

Figure 2: (a) Input Image; (b) Contrast enhancing (a) with Eq. (7); (c) Transformation (b) with Eq. (8); (d) Removing low-intensity pixels in (c) using Otsu thresholding algorithm; (e) Contrast enhancing (d) with Eq. (9); (f) Transformation of (e) with Eq. (10); (g) Removing low-intensity pixels in (f) using Otsu thresholding algorithm; (h) The detected fire region.

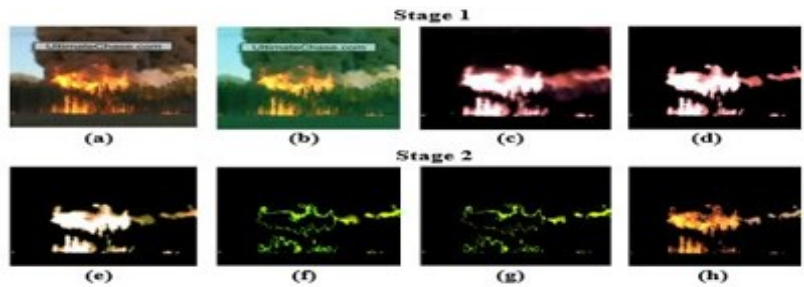

Figure 3: The binary image of (h) is a result of applying a logical OR operation on the binary image of (g) and the binary image that results from applying a threshold of 0.8 on each of the colour components of (d).

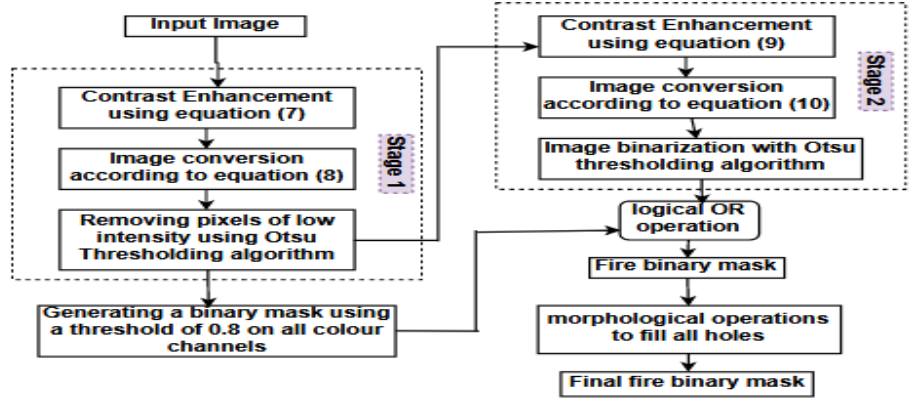

Figure 4: The proposed fire detection framework using a two-stage non-linear colour space transformation

### 2.4 Fire detection using the linear image transformation

This method involves a direct transformation of an image, using the proposed colour-differentiating conversion matrix. A gamma transformation of the input image is carried out using Eq. (9), followed by the transformation of the enhanced image according to Eq. (10). Lastly, the fire binary mask is created by applying Otsu thresholding algorithm o the converted image. Figure 5 shows the depiction of the steps involved in the fire detection algorithm.

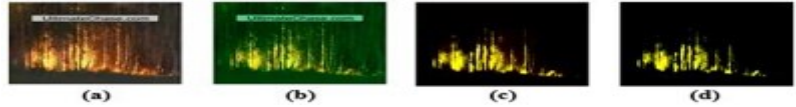

Figure 5. (a) Input Image; (b) Contrast enhance (a) with Eq. (9); (c) Transform (b) with Eq. (10); (d) Perform Otsu thresholding algorithm on (c).

## 3 Experimental Results

This section presents a detailed empirical assessment of our algorithm, using over 2000 still images extracted from 11 positive and negative video clips collected from the internet. The scenes in these videos include forest, car lights, daylight, buildings, and high colour similarity backgrounds. We manually label the ground truths for these frames and carry out a performance evaluation with alternative methods in the literature. Table 1 presents a qualitative comparison of our methods with those proposed in [12] - [14]. The results of this assessment show that our methods outperform the state-of-the-art algorithm especially on images of higher colour complexity. In Figure 6 and Figure 7, we use the false positive rate (FPR), false negative rate (FNR) and the F-score metrics to quantify the false alarm rate, missed detection rate and detection performance respectively. $FP$, $FN$, $TP$ and $TN$ are the false positive pixel counts, false negative pixel counts, true positive pixel counts and true negative pixel counts respectively. Movies with no fire scene are assessed with the FPR metric, and the fire videos are evaluated using the F-score metric. The proposed linear image transformation method reports the overall lowest false positive rate, followed by the non-linear image transformation method. However, the linear image transformation method has the overall worst missed detection or false negative rate, and sometimes fails to detect real fire pixels close to white colour. For example, it fails to detect fire in the video in row 11 of Table 1. In terms of F-score, the proposed non-linear image transformation method is the overall best performer, followed by the method in [12], and the third best performer is the linear image transformation method. However, these methods have a higher false negative rate than the methods in [13] and [14]. It is worth stating that most of these missed detections occur in the middle of fire regions which can easily be corrected using a morphological operation. The experimental results reveal the non-linear image transformation method has a better false positive rate than the method in [12] and a better false negative rate than the linear image transformation method.

| Videos | Khatami et al. [13] | Khatami et al. [14] | Khatami et al. [12] | The proposed linear image transformation | The proposed non-linear image transformation | Ground Truth images |
|---|---|---|---|---|---|---|
| 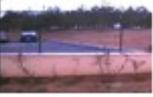 | 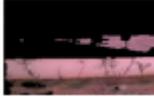 | 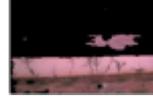 | 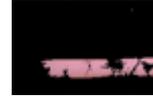 | 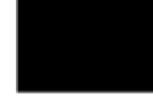 | 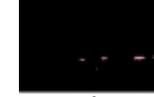 | 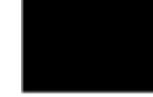 |
| 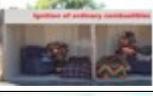 | 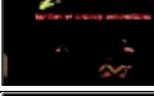 | 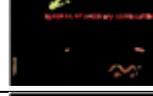 | 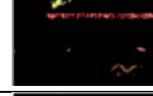 | 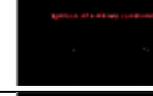 | 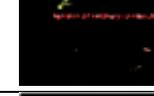 | 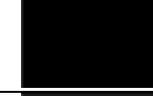 |
| 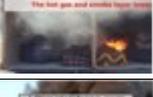 | 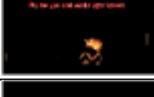 | 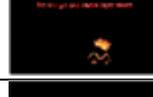 | 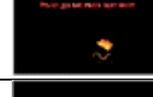 | 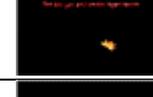 | 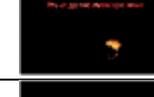 | 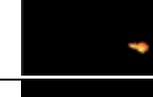 |
| 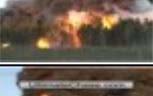 | 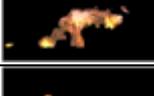 | 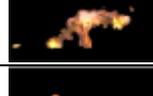 | 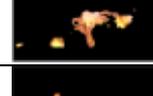 | 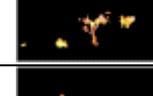 | 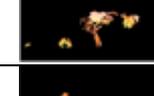 | 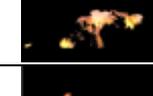 |
| 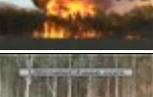 | 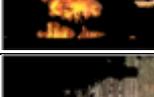 | 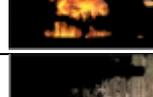 | 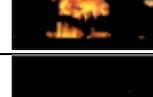 | 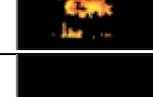 | 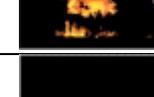 | 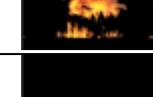 |
| 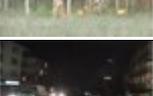 | 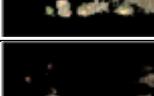 | 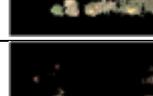 | 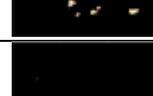 | 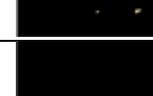 | 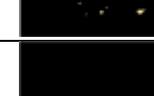 | 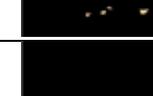 |
| 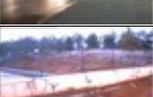 | 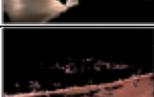 | 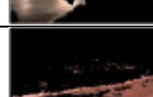 | 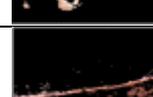 | 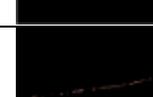 | 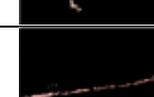 | 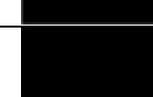 |
| 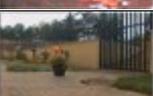 | 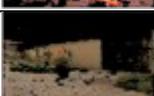 | 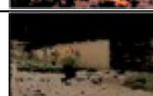 | 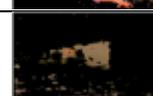 | 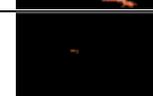 | 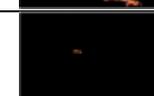 | 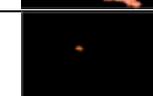 |
| 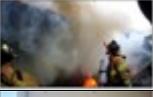 | 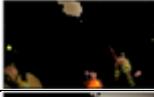 | 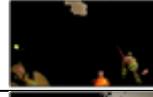 | 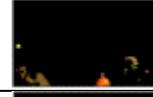 | 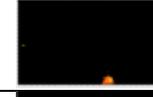 | 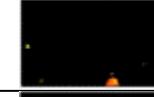 | 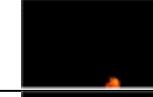 |
| 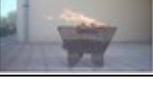 | 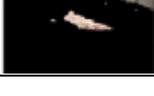 | 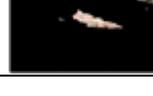 | 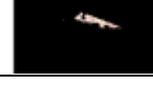 | 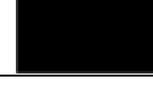 | 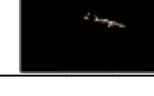 | 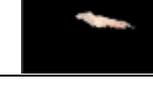 |

Table 1. Qualitative Comparison of the proposed methods with the state-of-the-art algorithms

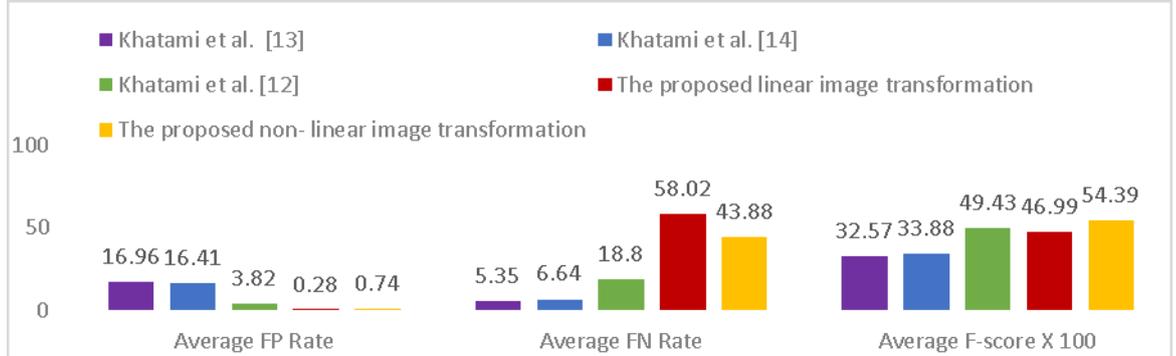

Figure 6: Evaluation of the proposed methods with the state-of-the-art methods on fire dataset.

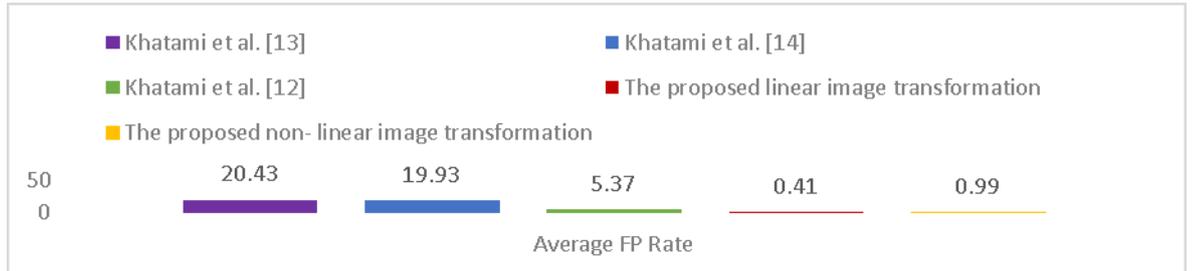

Figure 7: Evaluation of the proposed methods with the state-of-the-art methods on the non-fire dataset.

# 4 Conclusions

In this paper, we introduce a new colour-differentiating conversion matrix that is robust against false alarm. Subsequently, we use the proposed conversion matrix to design two fire detection frameworks. The first detection method is a two-stage, non-linear image transformation: In the first stage of transformation, we convert the contrast-enhanced input image with the conversion matrix proposed in Khatami et al. [12] and extracted the candidate fire pixels from the converted image, using Otsu thresholding algorithm. We further transform the image in the second stage, using the proposed conversion matrix, to reduce the possibility of a false alarm. The second method involves a direct transformation of the contrast-enhanced image with the proposed conversion matrix. The experimental result shows that the non-linear image transformation algorithm outperforms the state-of-the-art algorithm on the F score metric. The linear image transformation method reports the best false alarm rate, but on occasions fails to detect fire.

# References

[1] B. C. Ko, K. H. Cheong, and J. Y. Nam, "Fire detection based on vision sensor and support vector machines," *Fire Saf. J.*, vol. 44, no. 3, pp. 322–329, 2009.

[2] R. Chen, Y. Luo, and M. R. Alsharif, "Forest fire detection algorithm based on digital image," *J. Softw.*, vol. 8, no. 8, pp. 1897–1905, 2013.

[3] T. X. Truong and J. M. Kim, "Fire flame detection in video sequences using multi-stage pattern recognition techniques," *Eng. Appl. Artif. Intell.*, vol. 25, no. 7, pp. 1365–1372, 2012.

[4] T. H. Chen, P. H. Wu, and Y. C. Chiou, "An early fire-detection method based on image processing," in *Image Processing, 2004. ICIP'04. 2004 International Conference on*, 2004, vol. 3, pp. 1707–1710.

[5] W. B. Horng, J. W. Peng, and C. Y. Chen, "A new image-based real-time flame detection method using color analysis," in *2005 IEEE Networking, Sensing and Control, ICNSC2005 - Proceedings*, 2005, vol. 2005, pp. 100–105.

[6] T. Celik, H. Demirel, H. Ozkaramanli, and M. Uyguroglu, "Fire detection using statistical color model in video sequences," *J. Vis. Commun. Image Represent.*, vol. 18, no. 2, pp. 176–185, 2007.

[7] T. Celik, "Fast and efficient method for fire detection using image processing," *ETRI J.*, vol. 32, no. 6, pp. 881–890, 2010.

[8] T. Toulouse, L. Rossi, T. Celik, and M. Akhloufi, "Automatic fire pixel detection using image processing: a comparative analysis of rule-based and machine learning-based methods," *Signal, Image Video Process.*, vol. 10, no. 4, pp. 647–654, 2016.

[9] X. F. Han, J. S. Jin, M. J. Wang, W. Jiang, L. Gao, and L. P. Xiao, "Video fire detection based on Gaussian Mixture Model and multi-color features," *Signal, Image Video Process.*, vol. 11, no. 8, pp. 1419–1425, 2017.

[10] T. Celik and H. Demirel, "Fire detection in video sequences using a generic color model," *Fire Saf. J.*, vol. 44, no. 2, pp. 147–158, 2009.

[11] S. Y. Du and Z. G. Liu, "A comparative study of different color spaces in computer-vision-based flame detection," *Multimed. Tools Appl.*, vol. 75, no. 17, pp. 10291–10310, 2016.

[12] A. Khatami, S. Mirghasemi, A. Khosravi, C. P. Lim, and S. Nahavandi, "A new PSO-based approach to fire flame detection using K-Medoids clustering," *Expert Syst. Appl.*, vol. 68, pp. 69–80, 2017.

[13] Khatami, A., Mirghasemi, S., Khosravi, A., & Nahavandi, S. (2015a, July). An efficient hybrid algorithm for fire flame detection. In *Neural Networks (IJCNN), 2015 International Joint Conference on* (pp. 1-6). IEEE.

[14] A. Khatami, S. Mirghasemi, A. Khosravi, and S. Nahavandi, "A new color space based on k-medoids clustering for fire detection," in *Systems, Man, and Cybernetics (SMC), 2015 IEEE International Conference on*, 2015, pp. 2755–2760.

[15] R. Eberhart and J. Kennedy, "A new optimizer using particle swarm theory," in *Micro Machine and Human Science, 1995. MHS'95., Proceedings of the Sixth International Symposium on*, 1995, pp. 39–43.